\documentclass[12pt]{article}

\usepackage{graphicx,proof,color}

\usepackage[english]{babel}
\usepackage[T1]{fontenc}
\usepackage[utf8]{inputenc}

\begin{document}

\title{Explanation: from ethics to logic}
\author{Gilles Dowek\thanks{Inria, ENS-Paris Saclay, and CNPEN, {\tt gilles.dowek@ens-paris-saclay.fr}.}}
\date{}
\maketitle
\thispagestyle{empty}
\maketitle

\begin{abstract}
When a decision, such as the approval or denial of a bank loan, is
delegated to a computer, an explanation of that decision ought to be
given with it. This ethical need to explain the decisions leads to the
search for a formal definition of the notion of explanation. This
question meets older questions in logic regarding the explanatory
nature of proof.
\end{abstract}

Keywords: explanation, ethics, proof, cut, generalization

\section{Explanation: an ethical need}

When a person makes a decision regarding another person, for example
when a clerk refuses a bank loan to a customer, she ought to explain
her decision. Providing such an explanation realizes several values:
transparency, equality, agency, dignity...  Transparency, because the
customer wants to know the rules according to which her application
has been rejected. Equality, because she wants to make sure these
rules are the same for everyone. Agency, because providing such an
explanation enables the customer to improve her application. Dignity,
because providing an explanation sets up the customer as a rational
being, at the same level as the clerk.

If the clerk delegates this decision to a computer, such an
explanation of the decision must be generated together with the
decision itself.  Automatizing decision thus requires a formal
definition of the notion of explanation, while the interaction between
persons only required an informal one: a common agreement.

\section{Proof as an explanation}

\begin{figure}
  \begin{center}
    \includegraphics[height=4cm]{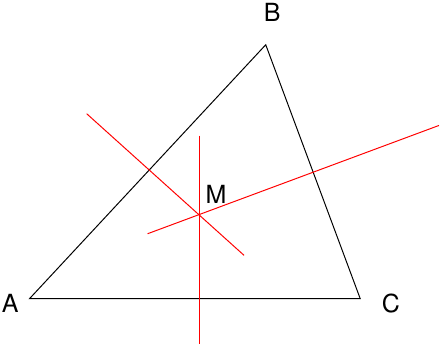}
  \end{center}
  \caption{If a point $M$ is on the perpendicular bisectors the segments
    $AB$ and $BC$, then $d(M,A) = d(M,B)$ and $d(M,B) = d(M,C)$, hence
    $d(M,A) = d(M,C)$ and $M$ is also on the perpendicular bisector of
    $AC$.\label{bisectors}}
\end{figure}

A first attempt is to define an explanation of a statement as a
logical proof of this statement.  For instance the proof, Figure
\ref{bisectors}, that the three perpendicular bisectors of the sides
of a triangle meet at one point, both shows that these bisectors meet
at one point, and also explains why.

The statement ``the perpendicular bisectors of the sides of the
triangle meet at one point'' is the thing that is explained and the
proof of this statement is the thing that explains it.

\section{Four proofs that are not explanations}

But, in a discussion, more epistemological than ethical
\cite{ComputationProofMachine}, we have already pointed that some
proofs do not seem to explain anything.

\subsection{The four color theorem}

\begin{figure}
  \begin{center}
    \includegraphics[height=4cm]{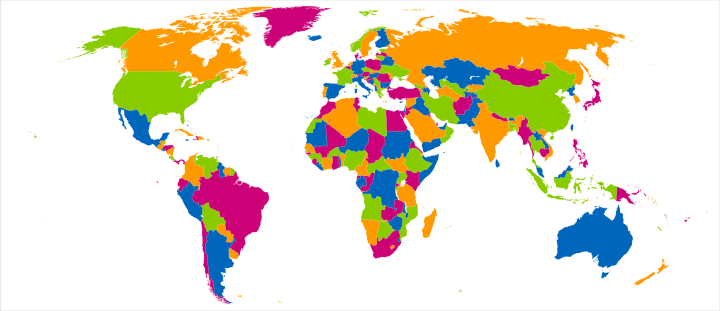}
   \end{center}
   \caption{Four colors are sufficient to color any map, in such a way
    that each country is of a color different from that of its
    neighbors. \label{fourcolor}}
\end{figure}

A first example is the proof of the four color theorem expressing that
every map is four colorable (Figure \ref{fourcolor}). This proof,
constructed in 1976 by Appel, Haken, and Koch
\cite{AppelHaken77,AppelHakenKoch77}, reduces the infinitude of all
possible maps to 1482 and checks, one by one, that these 1482 maps
verify some property, called {\em reducibility}.  This long case study
proves that the 1482 maps are all reducible, but it does not seem to
explain why.  Worse: this case study rather calls for an explanation,
than it provides one, as if we drew 1482 triangles and checked that
the perpendicular bisectors of the sides of each triangle met at one
point, we would rather think that such a coincidence needs to be
explained, than that it is, itself, an explanation.

\subsection{The weather forecast}

\begin{figure}
  \begin{center}    
    \includegraphics[height=5cm]{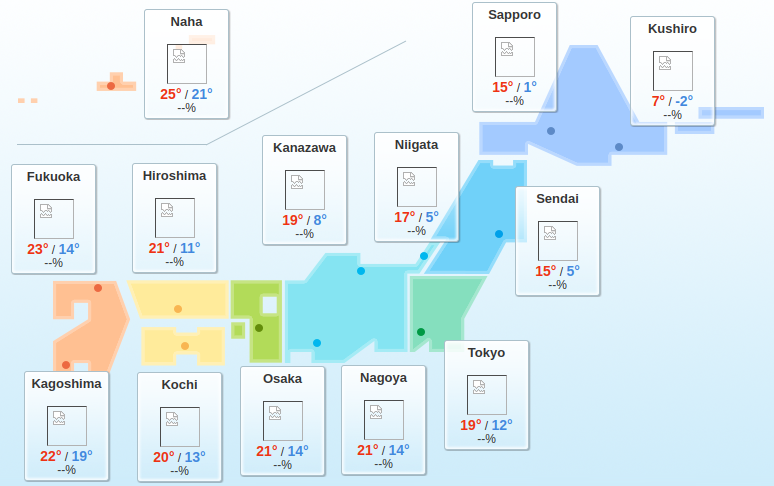}
  \end{center}
  \caption{Why $19^{\circ}\mbox{C}$ and not
    $18^{\circ}\mbox{C}$?\label{forecast}}
\end{figure}

Another example of a proof that does not seem to provide any
explanation is the computation of the weather forecast. For example,
if this computation forecasts a temperature of 19$^\circ$C in Tokyo
tomorrow (Figure \ref{forecast}), then we can view this computation as
a proof of the statement ``the temperature in Tokyo tomorrow will be
19$^\circ$C''.

Of course, unlike the judgement that the perpendicular bisectors of
the sides of the triangle meet at one point, that is analytic, the
judgement that the temperature in Tokyo tomorrow will be 19$^\circ$C
is synthetic, thus the statement ``the temperature in Tokyo tomorrow
will be 19$^\circ$C'' cannot be proved. So, this computation is a
proof that the temperature in Tokyo tomorrow will be 19$^\circ$C, not
in the real world, but in a fictive one, defined by the axioms of the
implemented numerical analysis model. That this fictive world
approximates the real world well enough is a falsifiable statement,
that can be temporarily supported by an empirical comparison of the
temperature series in the real world and in the fictive one.

This proof is not explanatory and the reason why it is not seems to be
that the numerical analysis computation leading to this forecast is
very long, even when using a massively parallel computer. Each step of
the computation seems to contain a bit of explanation, that we cannot
aggregate, when all these steps are put together.
  
\subsection{Data-centric algorithms}

Another example of a proof that does not seem to provide any
explanation is the use of data-centric algorithms, for instance
machine learning algorithms.

\begin{figure}
  \begin{center}
    \includegraphics[height=3cm]{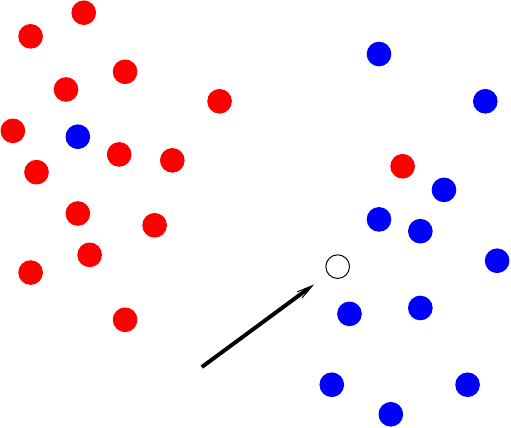}
  \end{center}
  \caption{{\color{red} Cats} and {\color{blue} dogs\label{datacentric}}}
\end{figure}

On Figure \ref{datacentric}, the points represent pictures in a
database.  The red points represent cat pictures and the blue ones dog
pictures.  The new point, in white, is, most likely, a dog picture as
its average distance to dog pictures is smaller than that to cat
pictures. But this proof of the statement ``The new picture most
likely represents a dog'' does not explain why this picture
represents, most likely, a dog, as would a proof analyzing the size of
the ears or the shape of the muzzle of the animal.

In this case, the reason why the proof is not explanatory seems to be
the large amount of data used by the algorithm.
Each piece of data seems to contain a bit of explanation, that we cannot
aggregate, when all these data are put together.

\subsection{The multiplication of arbitrary numbers}

A final example of a proof that does not seem to provide any
explanation is the multiplication on Figure \ref{multiplication}, of
two randomly chosen numbers. This multiplication proves that $7678
\times 3706 = 28 454 668$ but does not explain why this is the case.

A similar example is that of an algorithm that computes a
pseudo-random number: it would be non sense to attempt to explain why
this number is $28 454 668$.

These four examples: the four color theorem, the weather forecast, the
data-centric algorithms, and the multiplication, are four proofs that
do not seem to provide any explanation of the proved statement.  In
some cases, it is even difficult to imagine what an explanation would
look like: if we pick two arbitrary numbers and multiply them, what is
there to explain?

\begin{figure}
  \begin{center}
   \renewcommand{\tabcolsep}{1pt}
    \begin{tabular}{ccccccccccc}
      &&&&$7$&$6$&$7$&$8$\\
      &&&&$3$&$7$&$0$&$6$\\ \hline
      &&&$4$&$6$&$0$&$6$&$8$\\
      &&$0$&$0$&$0$&$0$&$0$&\\
      &$5$&$3$&$7$&$4$&$6$&&\\
      $2$&$3$&$0$&$3$&$4$&&&\\ \hline
      $2$&$8$&$4$&$5$&$4$&$6$&$6$&$8$\\
    \end{tabular}
  \end{center}
  \caption{A multiplication\label{multiplication}}
\end{figure}

\section{Another multiplication}

Unlike in that of Figure \ref{multiplication}, there seems to be
something to explain in the multiplication of Figure \ref{444}: one
can observe some regularity both in the first number to be multiplied:
$12345679$, although the $8$ seems to be missing, and in the result
$444444444$.  Thus, there seems to be something to explain: we can
seek for the reason why the result only contains the digit $4$.

\begin{figure}
  \begin{center}    
    \renewcommand{\tabcolsep}{1pt}
    \begin{tabular}{ccccccccccc}
      &$1$&$2$&$3$&$4$&$5$&$6$&$7$&$9$\\
      &&&&&&&$3$&$6$\\ \hline
      &$7$&$4$&$0$&$7$&$4$&$0$&$7$&$4$\\
      $3$&$7$&$0$&$3$&$7$&$0$&$3$&$7$&\\ \hline
      $4$&$4$&$4$&$4$&$4$&$4$&$4$&$4$&$4$\\
    \end{tabular}
  \end{center}
  \caption{If there were a 5 in the result, it would probably be a
    mistake.\label{444}}
\end{figure}

\subsection{A first bit of explanation}

Of course, a mathematically mature eye quickly notices that the second
number to be multiplied, $36$, has something to do with $4$: it is $9
\times 4$. So, the proved statement rephrases
$$12345679 \times 9 \times 4 = 111111111 \times 4$$
and this statement has another, more explanatory, proof than the
multiplication of Figure \ref{444}, obtained by first proving the
lemma
$$12345679 \times 9 = 111111111$$
and then multiplying both sides by $4$.

But, if both sides of this statement can be multiplied by $4$, they
can also be multiplied by any digit, for instance $7$, yielding the
statement
$$12345679 \times 63 = 777777777$$
Both theorems are consequences of a more general one
$$\forall n \in [1,9]~12345679 \times 9 \times n = 111111111 \times
n$$
and the explanatory nature of the second proof of the statement seems
to be a consequence of the fact that it proceeds by proving this
general result and then specializing it to $n = 4$.

This general result is the basis of a magic trick due to Lewis Carroll
\cite[p.78]{Gardner}: at a children's party, Lewis Carroll asked a
child to name her favorite digit, in his head he multiplied it by $9$
and asked the child to multiply $12345679$ by the product. She then
got a raw of nine repetitions of the named digit.

But this explanation is, of course, partial, as the statement
$$12345679 \times 9 = 111111111$$
still remains to be explained.

\subsection{A deeper explanation}

Instead of explaining the statement
$$12345679 \times 9 = 111111111$$
we can attempt to explain the equivalent one
$$111111111 / 9 = 12345679$$

If we perform the division of $111111111$ by $9$ (Figure
\ref{division}), we observe that the $n$-th digit of the result is
$n-1$, except the ninth one, which is a $9$ instead of an $8$.  But,
we can also observe that the $n$-th partial remainder is $n$, except
the ninth one, which is $0$ instead of $9$.  This can be proved by
induction on $n$: if the $n$-th partial remainder is $n$, then the $(n
+ 1)$-th partial dividend is $10 n + 1$, that is $9 n + (n + 1)$. Thus
the $(n + 1)$-th digit of the result is $n$ and $(n + 1)$-th partial
remainder $n + 1$.

\begin{figure}
  \begin{center}
    \renewcommand{\tabcolsep}{1pt}
    \begin{tabular}{ccccccccc|cccccccccccccccccccccccc}
      $1$&$1$&$1$&$1$&$1$&$1$&$1$&$1$&$1$&~&$9$\\
      ${\color{red} 1}$&$1$&&&&&&&&\multicolumn{11}{l}{--------------------}\\
      &${\color{red} 2}$&$1$&&&&&&&&$0$&$1$&$2$&$3$&$4$&$5$&$6$&$7$&$9$\\
      &&${\color{red} 3}$&$1$&&&&&&\\
      &&&${\color{red} 4}$&$1$&&&&&\\
      &&&&${\color{red} 5}$&$1$&&&&\\
      &&&&&${\color{red} 6}$&$1$&&&\\
      &&&&&&${\color{red} 7}$&$1$&&\\
      &&&&&&&${\color{red} 8}$&$1$\hspace*{1mm}&\\
      &&&&&&&&${\color{red} 0}$\hspace*{1mm}&\\
    \end{tabular}
  \end{center}
  \caption{The regularity is also in the partial remainders: 1, 2,
    3...\label{division}}
\end{figure}

But to deduce, from the fact that the $(n + 1)$-th partial dividend is
$9 n + (n + 1)$, the fact that the $(n + 1)$-th digit of the result is
$n$ and the $(n + 1)$-th partial remainder is $n + 1$, we need to use
the fact that $n + 1 < 9$.  This is the case until $n = 7$, but for $n
= 8$ the ninth partial dividend is $81$, that does not decompose into
$9 \times 8 + 9$, but into $9 \times 9 + 0$.  Thus the ninth digit of
the result is a $9$ and ninth partial remainder $0$.  It is thus a
good point to stop the division.

But this division can also be continued leading to generalize the
result: if instead of starting with a raw of nine digits $1$, we start
with a raw of $9 p$ digits $1$, we get a result which is a repetition,
$p$ times, of the sequence $012345679$. For example, with $p = 3$ we
get
$$111111111111111111111111111/9 = 12345679012345679012345679$$
hence
$$12345679012345679012345679 \times 9 = 111111111111111111111111111$$
and in general
$$(\sum_{i = 0}^{8} (8 - i) 10^i + 1) \times \sum_{j = 0}^{p - 1}
10^{9j} \times 9 = \sum_{k = 0}^{9 p - 1} 10^k$$ This generalization
yields another trick: ask the child to multiply the number
$12345679012345679012345679$ by the product and she will get a more
impressive sequence of twenty-seven repetitions of the named digit.

This remark can also be generalized to any base. For example, in base
$20$, with the digits $0, ..., 9, a, ..., j$, we have
$$123456789abcdefghj \times j = 1111111111111111111$$
---note that the $i$ is missing---and more generally, in any base
$b$\footnote{If the reader is not convinced by the proof based on the
  division algorithm, she can check that
$$(\sum_{i = 0}^{b - 2} (b - 2 - i) b^i + 1) \times (b - 1) = \sum_{i
    = 0}^{b - 2} (b - 2 - i) b^{i+1} - \sum_{i = 0}^{b - 2} (b - 2 -
  i) b^i + b - 1 = \sum_{i = 0}^{b - 2} b^i$$
and then multiply both sides by $\sum_{j = 0}^{p - 1} b^{(b - 1)j}$.}
$$(\sum_{i = 0}^{b - 2} (b - 2 - i) b^i + 1) \times \sum_{j = 0}^{p -
  1} b^{(b - 1)j} \times (b - 1) = \sum_{k = 0}^{(b - 1) p - 1} b^k$$
As the variables $p$ and $b$ can be any natural number, they are
implicitly universally quantified in this statement.

\begin{figure}
  \begin{center}    
    \renewcommand{\tabcolsep}{1pt}
    \begin{tabular}{ccccccccccccccccccccc}
      &$1$&$2$&$3$&$4$&$5$&$6$&$7$&$8$&$9$&$a$&$b$&$c$&$d$&$e$&$f$&$g$&$h$&$j$\\
      &&&&&&&&&&&&&&&&&$3$&$g$\\
      \hline
      &$h$&$e$&$b$&$8$&$5$&$1$&$i$&$f$&$c$&$9$&$6$&$2$&$j$&$g$&$d$&$a$&$7$&$4$\\
      $3$&$6$&$9$&$c$&$f$&$j$&$2$&$5$&$8$&$b$&$e$&$i$&$1$&$4$&$7$&$a$&$d$&$h$&\\
      \hline
      $4$&$4$&$4$&$4$&$4$&$4$&$4$&$4$&$4$&$4$&$4$&$4$&$4$&$4$&$4$&$4$&$4$&$4$&$4$\\
    \end{tabular}
  \end{center}
  \caption{At a vigesimal children's party.\label{vigesimal}}
\end{figure}

This generalization also yields a new magic trick: at a children's
party, in a vigesimal culture, ask a child to name her favorite digit,
in your head multiply it by $j$ and ask the child to multiply
$123456789abcdefghj$ by the product. She will get a raw of nineteen
repetitions of the named digit (Figure \ref{vigesimal}).

\section{Explanation: a definition} 

Instead of proving the statement
$$12345679 \times 36 = 444444444$$
with the multiplication algorithm, we first proved its
generalization
$$\forall n \in [1,9]~12345679 \times 9 \times n = 111111111 \times
n$$
in a generic way and then deduced the particular case
$$12345679 \times 36 = 444444444$$
As the quantification in the interval $[1,9]$ is finite, we could have
proved each of the nine cases, but this is not how we have proceeded:
we have first proved the statement
$$12345679 \times 9 = 111111111$$
and then multiplied both sides by $n$, in a generic way, that is without
enumerating the possible cases for $n$.

In the same way, instead of proving the statement
$$12345679 \times 9 = 111111111$$
using the multiplication or the division algorithm, we have proved
the more general statement 
$$(\sum_{i = 0}^{8} (8 - i) 10^i + 1) \times \sum_{j = 0}^{p - 1} 10^{9j} \times 9 =
\sum_{k = 0}^{9 p - 1} 10^k$$
or even 
$$(\sum_{i = 0}^{b - 2} (b - 2 - i) b^i + 1) \times \sum_{j = 0}^{p -
  1} b^{(b - 1)j} \times (b - 1) = \sum_{k = 0}^{(b - 1) p - 1} b^k$$
in a generic way---as the implicit universal quantification is on an
infinite domain, no enumeration is possible---and then deduced a
particular case for $p = 1$ and $b = 10$.

And the more general the intermediate statement, the more explanatory
the proof.

This remark leads us to define an explanation of a statement, not
merely as a proof of this statement, but as one that proves a general
statement in a generic way, and then specialize it to a particular
case.

This notion of a proof that proves a general statement in a generic
way, and then specializes it to a particular case already exists in
proof theory. A general statement is a statement of the form $\forall
x~A[x]$.  A generic proof of such a statement is a proof ending with
an introduction rule of the universal quantifier
$$\infer[\mbox{$\forall$-introduction}]
        {\forall x~A[x]}
        {\infer{A[x]}
               {\pi[x]}
        }$$
The specialisation of this proof to a particular case is the proof obtained
by adding an elimination rule of the universal quantifier
$$\infer[\mbox{$\forall$-elimination}]
        {A[t]}
        {\infer[\mbox{$\forall$-introduction}]
               {\forall x~A[x]}
               {\infer{A[x]}
                      {\pi[x]}
               }
        }$$
Such a proof ending with an introduction rule of the universal
quantifier followed by an elimination rule of this quantifier is
called {\em a cut on the universal quantifier}. It can be contrasted
with the proof $\pi[t]$ obtained by substituting the generic variable
$x$ with the particular term $t$, where the generality of the argument
is lost.

Thus, the definition above can be rephrased as the fact that an
explanation of a statement is a proof of this statement, that is a cut
on the universal quantifier. The proof of the statement
$$12345679 \times 36 = 444444444$$
that proceeds by proving the generalization 
$$\forall n \in [1,9]~12345679 \times 9 \times n = 111111111 \times
n$$
in a generic way and then deducing the particular case
$$12345679 \times 36 = 444444444$$
is a cut. That that proceeds by performing the multiplication is not.

This definition of the notion of explanation presupposes that the
statement to be explained can be formulated as a particular case of a
general statement. The discovery of this general statement is the
first step towards an explanation: the first step of the explanation
of the statement
$$12345679 \times 36 = 444444444$$
is to remark that $36$ and $444444444$ are both multiples of $4$ hence
that this statement can be rephrased
$$12345679 \times 9 \times 4 = 111111111 \times 4$$
and to formulate the hypothesis that, may be, the digit $4$ can be
replaced by any digit: may be, multiplying $12345679$ by the product
of any digit by $9$ yields a raw of nine repetitions of this digit.

\section{Revisiting the examples} 

We can now attempt to confront this definition to the six examples
discussed above: the bank loan application, the bisectors of a
triangle, the four color theorem, the weather forecast, the
data-centric algorithms, and the multiplication of two arbitrary
numbers. These examples fall in three categories.

In the first category, fall the examples for which we have an
explanation, that is a general statement and a generic proof of it. In
this category, falls the example of the bisectors of a triangle, for
which we first have a generic proof that the bisectors of the sides of
any triangles meet at one point, and then a specialization of this
proof to the considered triangle.

The example of the bank loan application also falls in this category,
when the clerk explains that all applications without a guarantor are
rejected. Indeed, in this case we first have a proof that of the
general statement ``all applications without a guarantor are
rejected'' and then a specialization of this proof to the considered
application. Often the proof of the general statement is a mere
reference to a rule of the bank, that can be considered as an axiom.

Note that, if the rule stated that the bank loan approval depends on
the gender, the sexual orientation, or the eyes color of the
applicant, it would also be a an explanation. And such a decision
would also be completely transparent. This shows that, from an ethical
point of view, the presence of an explanation alone, or transparency
alone, is not a sufficient condition for a good action. But the
presence of an explanation is a prerequisite to mobilize other values
such as gender equality.

In the second category, fall the examples for which we have a general
statement, but no a generic proof of it.  In this category, falls the
example of the four color theorem.  This theorem is a consequence of a
general statement: all maps belonging to a given set of 1482 maps are
reducible.  This statement is general, but the only known proof is to
enumerate the 1482 maps and check, one by one, that each of them is
reducible. This contrasts with the proof of the statement
$$\forall n \in [1,9]~12345679 \times 9 \times n = 111111111 \times
n$$
that first proves the statement
$$12345679 \times 9 = 111111111$$
and then generically multiplies the two sides by $n$.

Yet a proof with an analysis of 2 cases could still be called
explanatory. And a proof with a analysis of 12 cases may be called
less explanatory than one with 2 and more explanatory than one with
1482. Finding a threshold between explanatory proofs and non
explanatory ones may be as difficult as finding a threshold between
microscopic and macroscopic objects. Yet, this does not prevent a
quark to be microscopic and a whale to be macroscopic.

In the last category, fall the examples for which we do not even have
a general statement. Fall in this category the examples of the
weather forecast, the data-centric algorithms, and the multiplication
of two arbitrary numbers.  Because of this lack of a general
statement, we do not even see what an explanation would look like, we
do not even see what there is to explain.

In some situations, we try to find such a general statement. For
instance, there is a rule of thumb that, due to dominant west winds,
it is often the case that it rains in Strasbourg one day, if it rained
in Paris the day before. If this statement happened to be true, or
even statistically true, then it could be used as an explanation of
the weather in Strasbourg. Unfortunately, the forecast obtained by
using numerical analysis techniques is statistically more accurate.

In a similar way, if it happened to be true that a dog image more
likely represents a dalmatian than a dachshund, if it contains both a
lot of white pixels and a lot of black pixels. Then, this could be
used as an explanation of the fact that some image represents a
dalmatian. In machine learning, such an attempt to find general
statements is one possible direction towards explanatory artificial
intelligence.

Another example: take a database of landscape images, some of
them with a crescent moon and the others with a full moon. Use
this database to train a machine learning algorithm to identify if a
new image displays a crescent moon or a full moon. Then, pointing a
small square in the image displaying a crescent moon is considered
as a explanation of the statement ``The picture most likely represents
a landscape with a crescent moon''. Here again, pointing to this
square in the image is equivalent to formulate a general statement:
``All images containing this square, whatever the rest of the image
is, most likely represent a landscape with a crescent moon''.

\section{More examples of explanatory and non explanatory proofs: quantifier elimination}

\subsection{Explanatory proofs that Diophantine equations have no solutions}

The Diophantine equation
$$x^2 = 1800$$
has no solutions and the statement expressing it does not
$$\forall x~x^2 \neq 1800$$
has a simple proof: as $42^2 = 1764$ and $43^2 = 1849$, all numbers
smaller than or equal to 
$42$ are too small to be solutions
$$\forall x~(x \leq 42 \Rightarrow x^2 < 1800)$$
and all numbers larger than
$42$ are too large
$$\forall x~(x > 42 \Rightarrow x^2 > 1800)$$
Hence, using a case analysis, with two cases only, we get a proof of
the statement
$$\forall x~x^2 \neq 1800$$
This proof is explanatory as the proofs of
$$\forall x~(x \leq 42 \Rightarrow x^2 < 1800)$$
and of
$$\forall x~(x > 42 \Rightarrow x^2 > 1800)$$
based on the monotony of the square function, are generic.

\subsection{Non explanatory proofs that Diophantine equations have no solutions}

The statement
$$\forall x~(x \leq 42 \Rightarrow x^2 < 1800)$$
unlike
$$\forall x~(x > 42 \Rightarrow x^2 > 1800)$$
uses a quantification over a finite domain. Hence it can also be proved
by a simple enumerations of the $43$ cases $x = 0$, $x = 1$, ..., $x = 42$.

Because of this enumeration, the obtained proof is not explanatory
anymore. But this method can be generalized to any univariate
Diophantine equation
$$a_n x^n + a_{n-1} x^{n-1} + ... + a_1 x + a_0 = 0$$
When $x$ is larger than  
$b = n \max(|a_0|, ..., |a_{n-1}|)$, then $|a_n x^n|$ is larger than
$|a_{n-1} x^{n-1} + ... + a_1 x + a_0|$ and $x$ thus cannot be a solution.
Therefore, the statement
$$\exists x~(a_n x^n + a_{n-1} x^{n-1} + ... + a_1 x + a_0 = 0)$$
is equivalent to
$$\exists x~(x \leq b \wedge a_n x^n + a_{n-1} x^{n-1} + ... + a_1 x + a_0 = 0)$$
where the infinitude of all possible natural numbers has been reduced
to $b + 1$, that can be checked one by one.

This method called {\em quantifier elimination} because the
existential quantifier is replaced with a bounded one and then with an
enumeration, always produces non explanatory proofs.

Besides the non existence of solutions of univariate Diophantine
equations, quantifier elimination is used in the proof of the four
color theorem, and in the proofs built with Presburger's, Skolem's,
and Tarski's algorithms.

\section{From proofs to algorithms}

In contemporary logic, the notion of proof is not a primitive notion
anymore, but after Brouwer, Heyting, and Kolmogorov, it has been
defined in terms of a more primitive one: that of algorithm.

In particular a proof of a statement of the form $\forall x~A[x]$ is
an algorithm mapping every term $u$ to a proof of $A[u]$.  The proof
$$\infer[\mbox{$\forall$-introduction}]
        {\forall x~A[x]}
        {\infer{A[x]}
               {\pi[x]}
        }$$
is the algorithm mapping the term $u$ to the proof $\pi[u]$.

The explanation of $A[t]$ 
$$\infer[\mbox{$\forall$-elimination}]
        {A[t]}
        {\infer[\mbox{$\forall$-introduction}]
               {\forall x~A[x]}
               {\infer{A[x]}
                      {\pi[x]}
               }
        }$$
thus contains an algorithm: that mapping the term $u$ to the proof
$\pi[u]$, and an input value for this algorithm: the term $t$.

This leads to define, more generally, an explanation as a pair formed
with an algorithm and an input value for this algorithm. When this
algorithm is applied to this value, it returns a proof of the
explained statement.

According to Brouwer, Heyting, and Kolmogorov, a proof of a statement
of the form
$$\forall x~(A[x] \Rightarrow \exists y~B[y])$$
is an algorithm mapping each pair formed with a term $t$ and a proof
of the statement $A[t]$ to a pair formed with a term $u$ and a proof
of the statement $B[u]$.  If the proofs of the statements of the
form $A[.]$ and $B[.]$ are trivial, for example if these statements
belong to a decidable fragment or if they can be directly observed,
then a proof of this statement is simply an algorithm mapping each
term $t$ such that $A[t]$ to a term $u$ such that $B[u]$, this term
$u$ playing the r\^ole of a proof of $\exists x~B[x]$.  Then, an
explanation of the statement $\exists y~B[y]$ is simply a pair formed
with
\begin{itemize}
\item an algorithm mapping each term $t$ such that $A[t]$ to a term
  $u$ such that $B[u]$
\item and an input value for this algorithm, that is a term $t$ such
  that $A[t]$.
\end{itemize}

This notion gets closer to the usual notion of explanation.  If $A[x]$
is the statement ``the order $x$ has been placed yesterday'' and
$B[y]$ is the statement ``the book $y$ has been delivered by the
postman today''.  Then, an explanation of the statement $\exists
y~B[y]$: ``a book has been delivered by the postman today'' is a pair
formed with an algorithm mapping orders placed to books
delivered---such an algorithm may be called a {\em book shop}---and an
order that has been placed: a book has been delivered today because an
order has been placed yesterday at the book shop.

Note that the book itself is a proof of the statement ``a book has
been delivered'', but it is not an explanation of this statement. In
contrast, the pair formed with the book shop and the order placed is
such an explanation.

\section{The weather forecast paradox}

We may wonder if this definition of an explanation as a pair formed
with an algorithm and an input value for this algorithm is not too
general.  In particular, as the weather forecast of Figure
\ref{forecast} is obtained by applying an algorithm performing
numerical analysis computation to sensor data, the pair formed with
the algorithm and the sensor data would, according to our definition,
be an explanation of the forecast, even if we do not see a general
statement, this forecast would be a particular case of.

But, as already remarked, this proof is not explanatory, not because
it is not formed with an algorithm and an input value for this algorithm,
but because the program expressing this algorithm is large and the
execution of this algorithm is long.  So, nobody can trace the
computation step by step.  This situation can be contrasted with that
of the explanation of the statement
$$12345679 \times 9 \times 4 = 111111111 \times 4$$
or even the statement 
$$12345679 \times 36 = 444444444$$
that is formed with a proof of the statement 
$$\forall n \in [1,9]~12345679 \times 9 \times n = 111111111 \times n$$
and the input value $4$. In this case, the algorithm is expressed by a
short program and its execution is fast.

This situation is common in ethics: if a clerk refuses a bank loan to
a customer, not only she ought to explain her decision, but she ought
to explain her decision in a way that can be understood by the
customer, for example, in a language that the customer speaks. This is
why, for example, if a person is sued in a country she does not
understand the language of, the tribunal ought to provide an
interpreter.

This means that the quality of the explanation does not only depend on
the width of the possible input values of the algorithm: the
generality of the statement, but also on the small size of the program
expressing the algorithm and the low complexity, that is the small
execution time, of this algorithm. As these notions---small,
fast---are quantitative rather than qualitative, the notion of
explanation also is.

So it is possible to order explanations: the shorter the program and
the faster its execution, the deeper the explanation. Yet, it is also
possible to identify thresholds. For instance, using the notion of 
Kolmogorov complexity, we can call the pair formed with an algorithm
and an input value an explanation if the size of the program
expressing the algorithm and of the data expressing the input value is
smaller than the size of the explained statement.

\section{From proof to explanation}

Checking the 1482 cases of the four color theorem, applying the finite
element method, using a data-centric algorithm, checking the statement
$$\forall x~(x \leq 42 \Rightarrow x^2 \neq 1800)$$
or using Tarski's algorithm with a pencil and a paper, and even with
a pocket calculator, is tedious, and sometimes impossible.  The size of
the data, the size of the program expressing the algorithm, the
complexity of the algorithm... require the use of a computer.

The same size of data, size of the program, and complexity of this
algorithm make the proof non explanatory.  This explains that non
explanatory proofs have been built mostly after the invention of the
computer.

The use of computers both made possible the construction of many non
explanatory proofs, and the need of a formal definition of the notion
of explanation.

We propose here a first attempt towards such a definition: a pair
formed with a small and fast algorithm and an input value to this
algorithm such that the algorithm applied to the value produces a
proof of the statement to be explained.  So, the articulation between
proof and explanation and that between computability and complexity
are parallel.

Philosophy of computer science first focused on epistemological
questions, then ethical questions were also raised. The notion of
proof---whether explanatory or not---was the right tool to address
many epistemological questions, but it seems that the notion of
explanation is needed to address the ethical ones.

Thus, the articulations between epistemology and ethics, between
proof and explanation, and between computability and complexity
are parallel.

\end{document}